\documentclass{emulateapj}

\usepackage[dvips,usenames]{color}
\usepackage{natbib}
\usepackage{amsmath}
\usepackage{amssymb}
\usepackage{graphicx}
\usepackage{dcolumn}
\usepackage{rotating}
\usepackage{longtable}
\usepackage{pstricks}
\usepackage{fix-cm}

\newcommand\T{\rule{0pt}{3.5ex}}

\newcommand\B{\rule[-2.1ex]{0pt}{0pt}}

\setkeys{Gin}{width=0.75\textwidth}
\graphicspath{{./}{figures/}}

\shorttitle{Viscoelastic Tidal Exoplanets}
\shortauthors{Henning et al.}

\begin{document}

\title{Tidally Heated Terrestrial Exoplanets: \\
Viscoelastic Response Models}

\author{Wade G. Henning\altaffilmark{1}}
\affil{Earth and Planetary Science Deptartment, Harvard University, 20  
Oxford Street, Cambridge, MA 02138, USA}
\email{henning@fas.harvard.edu}

\author{Richard J. O'Connell}
\affil{Earth and Planetary Science Dept., Harvard University, 20  
Oxford Street, Cambridge, MA 02138, USA}

\and

\author{Dimitar D. Sasselov}
\affil{Harvard-Smithsonian Center for Astrophysics,
60 Garden Street, Cambridge, MA 02138, USA}

\altaffiltext{1}{Corresponding author}

\begin{abstract}
Tidal friction in exoplanet systems, driven by orbits that allow for durable nonzero eccentricities at short heliocentric periods, can generate internal heating far in excess of the conditions observed in our own solar system. Secular perturbations or a notional 2:1 resonance between a Hot Earth and Hot Jupiter can be used as a baseline to consider the thermal evolution of convecting bodies subject to strong viscoelastic tidal heating. We compare results first from simple models using a fixed Quality factor and Love number, and then for three different viscoelastic rheologies: the Maxwell body, the Standard Anelastic Solid, and the Burgers body. The SAS and Burgers models are shown to alter the potential for extreme tidal heating by introducing the possibility of new equilibria and multiple response peaks. We find that tidal heating tends to exceed radionuclide heating at periods below 10-30 days, and exceed insolation only below 1-2 days. Extreme cases produce enough tidal heat to initiate global-scale partial melting, and an analysis of tidal limiting mechanisms such as advective cooling for earthlike planets is discussed. To explore long term behaviors, we map equilibria points between convective heat loss and tidal heat input as functions of eccentricity. For the periods and magnitudes discussed, we show that tidal heating, if significant, is generally detrimental to the width of habitable zones. 

\end{abstract}


\section{Introduction}
\label{intro}
  
The discovery and study of planetary systems around different stars has revealed a rich diversity of orbital architectures, many of them not anticipated. Among the surprises is wide evidence for planet migration, orbital resonances, and the ubiquity of high orbital eccentricities. These interactions may often lead to orbits that allow for durable nonzero eccentricities close to the star, especially for terrestrial mass planets. Our goal is to investigate the range of tidal magnitudes that result from such orbital conditions. This paper examines the global temperature behavior of a simplified terrestrial planet during long-term extreme tidal heating, perhaps driven by mean motion or secular orbital resonances, using several different models of viscoelastic material response.

We first present results across a range of orbit periods using a blackbody model with fixed material parameters. This method suggests that at eccentricities of around 0.1, tidal heating becomes globally significant generally at heliocentric periods below 10-30 days, and large enough to drive global-scale partial melting at periods near 1-2 days. We map this range of extreme tidal activity in more detail using viscoelastic methods, and by modeling both temperature-dependent viscosity and melting. We investigate tidal work functions for three anelastic rheological rock models: the Maxwell, Standard Anelastic Solid (SAS), and Burgers bodies \citep{NowickBerry1972, Sabadini1987, Cooper2002}. Each of these alternative rock models have the potential to exhibit complex behaviors because their frictional work function is non-monotonic in temperature. Lastly, behaviors and extreme tidal equilibrium states are explored by linking heat input to a parameterized convection model.    

In this paper we use the term supertidal to refer to planets where tidal heating contributes significantly to the global heat budget. While the conditions needed for enduring extreme tides are expected to be rare, we are motivated by the rapid expansion of exoplanet discoveries. Transiting searches such as Kepler \citep{Borucki2008}, COROT \citep{Borde2003} and MEarth \citep{Charbonneau2008} all have the potential to discover a terrestrial planet of this supertidal type. A Hot Earth trapped in 2:1 resonance with a Hot Jupiter may also be detectable via careful transit timing \citep{Holman2004, MillerRicci2008}, however no such timing patterns have yet been found.  

We focus here on hot planetary cases, where tidal heat competes with insolation for significance. This is as opposed to cold exomoons \citep{Scharf2006}, where it is easier for tidal heating to play a more dominant role, but detectability may be further off. Tidal heat may also play a role in enlarging the radii of some gas giants \citep{Bodenheimer2003}, however we focus on rocky objects. Study of tidal exoplanets is important partly because eccentricities in the overall exoplanet population are higher than expected \citep{Butler2006, Tremaine2004, Namouni2005}, e.g. GJ436b at e=0.15 \citep{Deming2007}, suggesting the importance and prevalence of multiple planet interactions. 

Heating predictions for Hot Earths can become quite large, on the order of millions of terawatts. However global-scale partial melting of a mantle can begin with only a few tens of terawatts of added tidal heat. As on Io, widescale melting lowers average viscosities and can lead to tidal shutdown. We discuss this shutdown behavior and how it can be altered by the high pressures in an Earth-mass body.

Our methods for a Hot Earth follow those used to analyze the anomalously high heating of Io \citep{Smith1979, Carr1998, McEwen2000} and to a lesser extent Europa \citep{Cassen1979, Squyres1983, Ojakangas1989} and Enceladus \citep{Porco2006}. These bodies have been the subject of intense analysis in terms of tidal heating via constant parameters \citep{Reynolds1980, YoderPeale1981, Tackley2000}, viscoelastic parameters and layered deformation \citep{Segatz1988, Takeuchi1962}, advection \citep{Moore2001, Monnereau2002}, and thermal-orbital coupling through orbital eccentricity \citep{HussmannSpohn2004, Showman1996, PealeLee2002, Showman2003}. Here we consider melting and tides as a function of temperature and forcing frequency, but do not explicitly model orbital feedback or include variations by layer, latitude or longitude.

In section 2 we discuss orbital configurations that can give rise to extreme tidal heating. In section 3 we present the model and results for a fixed Quality factor ($Q$) and Love number ($k_2$) approach. In section 4 we discuss the derivation of complex rigidities, Love numbers, and work functions for three viscoelastic models and show results in section 5. In section 6 we discuss long term behaviors, tidal equilibria, and present bifurcation diagrams for the movement of tidal-convective equilibria points through changes in tidal forcing. Lastly, in section 7, we address issues such as advective heat transport, magma ocean production, and habitability. 

\section{Resonance and Stability}
\label{resonance}

The first condition for extreme tidal heating is proximity to a massive host, providing a large change in gravity gradient between pericenter and apocenter. While moons often meet this criterion, only now have a large number of planets been detected in regions near stars where tidal heating becomes of geological concern. For extreme tides around a typical main sequence star, planets must be well inside the 88 day orbit of Mercury, but precisely within the 1-20 day range of Hot and Warm Jupiters. An apparent number density peak of Hot Jupiters exists near $\sim$4 day periods, and thus a mirrored peak in 2:1 resonant Hot Earths could occur near $\sim$2 day orbits (unless proximity to the star causes better stability only further away).

The second condition for tidal heating is an elliptical orbit maintained by some durable perturbation, such as the regular effect of Europa on Io. Orbital resonances, and in particular the 2:1 mean motion resonance, provide an effective mechanism to support long-term tidal heat production. Short-term tides from obliquity and non-synchronous spin damp quickly, leaving the body briefly warmed, but thereafter tidally quiescent. Nonzero eccentricities can be more enduring, and can arise from mean motion resonances, secular perturbations, secular resonances, or recent interactions.

\subsection {Mean Motion Resonances}

Several mean motion resonances (MMR) have been observed in extrasolar planet systems \citep{LeePeale2002, Gu2003, Kley2005}. For several reasons, the 2:1 MMR and to a lesser degree the 3:2 MMR are most favorable for exciting a terrestrial planet to extreme long-term tides. In a resonance $(p+q):p$, the order $q$, determines how often orbital conjunctions are co-located (e.g. at pericenter). Frequent coherent conjunctions in the 2:1 and 3:2 cases increase resonance perturbation strength relative to higher orders such as the 3:1 or 5:2 resonances. The 2:1 MMR can be particularly stable because of its large physical gap in semi-major axis from other resonances, whereas the 4:5 and higher MMRs can become packed together in real space, leading to chaos and ejections.  

The large resonance widths of the 2:1 and 3:2 MMRs also favor longer durations for extreme tides. Since tidal heating will drive a system away from a state of perfect resonance, wider cases can induce greater heating prior to resonance breaking.

The order of capture for a migrating perturber favors the presence of bodies in the 2:1 and 3:2 resonances. Capture here is likely, and bodies must have either missed capture or have been otherwise scattered to reach and populate the resonances closer to the perturber. 

High perturber eccentricity can be both favorable and detrimental for extreme tidal forcing. High perturber eccentricity in the 2:1 resonance may increase the amplitude of inner body eccentricity forcing. However, if pumping is too great, orbit crossings can lead to ejections and clearing. 

\citet{Haghighipour2007} and \citet{RiveraHaghighipour2007} analyze the stability of terrestrial planets in or near resonances with short-period gas giants and known exoplanets and identify multiple orbits both inboard and outboard of the perturbing bodies as stable or quasi-stable.

Once strong tides are established, they tend to force the system away from perfect resonance. \citet{Terquem2007} analyze the dynamics of short period Hot Earths and find tides eventually break exact 2:1 resonances, but remain in near-resonance with nonzero eccentricities during later stage tidal dissipation. Disk turbulence may also break MMRs \citep{Lecoanet2009} but leave bodies in a near-commensurable state vulnerable to ejections. 

\subsection {Secular Perturbations}

Secular perturbations occur in two-planet systems and lead to an equilibrium forced eccentricity. In this case there is no resonance to be broken, so such forced eccentricities can provide a very long term source for tides \citep{Mardling2006, Zhou2008}. The effective strength of this perturbation against damping depends on the secular forcing timescale (a function of the planets' sizes and locations) and the tidal dissipation timescale (a function of heating magnitude).  

Following \citet{MardlingLin2004}, the secular equilibrium eccentricity for an interior Hot Earth of semi-major axis $a_{sec}$ perturbed by an outer Jupiter (not in 2:1 resonance), is $\sim$6\% of the eccentricity of the perturber for $a_{pert}/a_{sec}$=10, and 19\% for $a_{pert}/a_{sec}$=30. Relativistic corrections can reduce these values somewhat further. Secular timescales are often of order 10,000 years, while moderate tidal timescales are often on the order of a few million years, suggesting that conditions can exist where modest geologically significant tidal heating is supported by secular perturbations alone. While extreme tides may damp forced eccentricities, temporarily cessations perhaps due to mantle melting will allow windows for secular forced eccentricities to be restored. 

Secular resonance occurs in multi-body systems. Unstable secular resonances occur at fixed periods, which if a small planet crosses can lead to high eccentricities and ejection. In our solar system these unstable points lie due to Jupiter and Saturn near 0.5, 2, 12.5, and 17.5 AU \citep{MurrayDermott2005}. Outside these singularities, secular resonance drives eccentricities of order 0.0-0.06. Unlike with MMR cases, there is no locking mechanism, so a planet under the influence of a secular resonance can migrate inwards due to tidal heating. If in one of the singularities, migration will be more rapid, leading to a brief episode of increased heat output. This effect will be magnified the closer to a star it occurs. Such secular singularity crossings are likely explanations for sudden bursts of eccentricity, but not of sustained extreme heating.  

\subsection {Capture}

Mean motion resonances in exoplanet systems are thought to arise from various convergent migrations in the late stages of planet formation \citep{Lee2004}. Differential inwards migration is common in early solar systems \citep{Thommes2005} and planetary moon systems \citep{Kley1999, CanupWard2002}, and is a possible origin of the 1:2:4 resonant Laplace relation among the Jovian satellites \citep{PealeLee2002}. Resonance capture occurs in convergent cases when eccentricities are below a critical threshold. Capture does not occur in divergent cases. For large initial eccentricities, capture becomes probabilistic, with a known, albeit complex, expression for $P_{capture}$. Following \citet{MurrayDermott2005}, the critical eccentricity for internal 2:1 resonance capture in an Earth-Jupiter-Sol system is 0.152, and for a 3:2 capture is 0.122. Threshold values can be higher for M dwarf and super-Jupiter cases. A terrestrial planet's effective starting eccentricity is a function of both is secular forcing environment and its impact history. While tides are unlikely to cause much damping at 1AU during the chaotic growth phase, many early inner planets may have $e<0.15$ and be candidates for capture. 

One scenario to form a supertidal Hot Earth is for a migrating future Hot Jupiter to capture and sweep an inner rocky planet along in its 2:1 inner resonance \citep{YuTremaine2001}. For young solar systems, nebular disk torques are expected to dominate and induce Type I and Type II migrations \citep{Thommes2005}. After a gas giant clears a gap in its local region, Type II migration takes hold, and is considered the source of Hot Jupiters. Terrestrial planets may experience modest random-walk style Type I migration \citep{Laughlin2004}, but remain generally in place, setting up a convergent pattern. Sweeping capture may collect multiple bodies from an inner solar system into the same resonance. The sweeping merger of multiple large embryos would lead towards a super-Earth mass with an initial hot start.  

It is favorable if the MMR reaches a rocky planet prior to a secular resonance that could pump up high eccentricities. Secular resonance positions depend on the overall solar system configuration and shift when precession rates or masses change \citep{Nagasawa2005}. While many scenario geometries can prevent a secular singularity from disturbing a candidate 2:1 Hot Earth, one such case is a system with only one gas giant and thus a lack of unstable secular resonances. Since only 5\% of sunlike stars appear to have gas giants \citep{UdrySantos2007}, such cases may be common.
 
\citet{YuTremaine2001} and \citet{Lee2004} analyze the resonant sweeping of exoplanets in detail. As a giant outer planet continues to migrate, it progressively increases the trapped inner body's eccentricity, and later inclination. After migration by a factor of 4 in semi-major axis (for a 2:1 case), the inner body eccentricity is large enough to cause close encounters with the host star. Loss to the star can occur, or perhaps tidal damping and disk interactions can hold eccentricity to an equilibrium value as migration proceeds further. If the trapped Hot Earth is able to survive in resonance all the way down to short periods, it may do so with a large initial reservoir of eccentricity to feed further tides. Undamped resonant migration by a factor of 9 leads inner bodies to resonance release on nearly circular retrograde orbits. For these reasons 2:1 trapped Hot Earths may be more favorable at dimmer stars, where snow lines are closer and requisite migration distances shorter.   
	 	 
Alternately, orbital resonances can be traps where scattered planetesimals congregate without invoking the sweeping capture mechanism above. \citet{Mandell2007} demonstrate in numerical simulations of gas-disk induced migrations the formation of a variety of Hot Earths via scattering, often near the 2:1 resonance points of migrating Hot Jupiters, and with inner solar systems often cleared of further material. Denser inner gas disks appear correlated with having Hot Earths at the end of their 200 Myr simulations. These simulations did not include tidal damping or attempt to address long term stability.     

Outward migration due to the influence of a tidal bulge on a fast rotating primary is considered necessary for the Galilean system to remain in equilibrium \citep{PealeLee2002}. Analogous behavior may occur for hot exoplanets. For the outward migration of a single body not in a resonance, eccentricity will simply increase. High $e$ in moon systems is considered a signature of outward tidal evolution in older systems. When the joint migration of a 2:1 resonant system is dominated by dissipation in the primary, eccentricities will be limited by $Q$ of the primary, which sets an effective limit on the dissipation in both secondaries \citep{dePaterLissauer2005}. For our analysis we make no assumptions about tides in the star, nor the star's rotation rate relative to its inner planets. We generally consider $e$ as a free parameter and compute resulting tidal heat rates.  

Full treatment of tides requires thermal-orbital coupling \citep{HussmannSpohn2004}, which this paper does not model. In resonant systems this coupling is complex, and extreme friction can lead to evolution away from exact resonance and possible resonance breaking. In non-resonant systems tides simply lead to circularization and semi-major axis drift.

\subsection {Circularization}
\label{circularization}

While ongoing perturbations are favorable to supertidal conditions, they are not necessary. Circularization timescales may still be of the order 0.1-10 Gyr for short period Earth-mass planets. \citet{Jackson2008} also examine the tidal heating of non-resonant terrestrial exoplanets, and discuss how tidal orbital migration further lengthens circularization times.   

Often circularization times are reported without regard to a planet's internal response, leading to results that may be too short. For example a Hot Earth in a 2 day orbit of a 1 solar mass host with 1$\times$10$^4$ terawatts (TW) of dissipation at $e=0.1$ will have a 5.4 billion year damping timescale, while at 1$\times$10$^6$ TW circularization takes only 54 million years. The higher dissipation rate however may be unsustainable for a rocky body due to global onset melting (or more generally viscoelastic de-tuning).  Instead, a migrating planet may briefly experience a few million TW in tides, then rapidly warm to a small partial melt fraction. Decreasing viscosities thereafter produce only a few thousand TW of tidal heat, which may then persist for a solar system lifetime even without support from a resonance. This phenomenon is discussed further in section \ref{equilibria}. It is important when calculating circularization timescales using a fixed Quality factor $Q$ to take into consideration whether the planet itself can geologically produce the subsequent rate of dissipation.

Countless complications to the arguments in this section exist: High exoplanet eccentricities could in part be caused by unseen high inclination perturbers, as occurs via the Kozai mechanism \citep{TakedaRasio2005}. Nonzero secular precessions and inclinations mix with mean motions to obtain a variety of higher order resonances. At short periods, a star's oblateness, tidal bulge, atmosphere, and relativistic effects may also influence stability. 

Observations will ultimately decide the matter. Overall we consider it likely enough that some terrestrial planets can be swept or scattered into resonances by migrating Hot Jupiters, or may otherwise have their eccentricities sustained at nonzero values for geologically significant times, to move forward and consider the tidal heat magnitudes that then result. 

\section{Fixed Q Tidal Model}
\label{fixedQ}

Tidal heating is modeled in many ways, but the starting point is the global heat generation rate $\dot{E}_{tidal}$ \citep{PealeCassen1978, Peale1979, Showman1996}. For a homogeneous spin-synchronous body whose stiffness and viscous dissipation are both assumed to be constant and uniform, the global tidal heat rate can be expressed following the detailed derivation in \citet{MurrayDermott2005}. 
	
\begin{equation}
\label{SnMEdot}
\dot{E}_{tidal}=\frac{21}{2} \frac{k_2}{Q}
\frac{G M_{pri}^2 R_{sec}^5 n e^2}{a^6}
\end {equation}

$\dot{E}_{tidal}$: Tidal heat production rate, $watts$

$a$: Semi-major axis

$e$: Eccentricity 

$G$: The gravitational constant

$M_{pri}$: Mass of the primary\footnote {Because subscripts such as $M_s$ and $M_p$ can be confused between planet, satellite, sun, primary and secondary, we adopt three letter subscripts and use primary to refer to the central mass, whether it is a star or planet, and consider tides generated inside an orbiting secondary.}

$R_{sec}$: Radius of the secondary

$k_2$: Second-order Love number of the secondary 

$Q$: Quality factor of the secondary
\\

The masses, radii, and orbital parameters $a$ and $e$ for exoplanets have the potential to be directly measured. In contrast, $Q$ and $k_2$ must be inferred, and represent in only two (albeit convenient) lumped scalar terms all the complex internal material properties of a deforming planet.  After the high power terms in equation \ref{SnMEdot} are known, almost all uncertainty in the tidal heating comes from $k_2$ and $Q$. As discussed in section \ref{viscosection}, these terms together are actually a function of both forcing frequency and temperature. But first we can gain insight by demonstrating results of equation \ref{SnMEdot} simply by picking commonly invoked values for $k_2$ and $Q$.

The Quality Factor $Q$ is an inverse damping term characterizing energy lost to friction, and can be variously defined for a material, mode, process, or compound process. Following \citet{OconnellBudiansky1978}, the Q for an oscillator may be written:

\begin{equation}
Q(\omega) = 2\pi \frac{(2E_{ave}(\omega))} {\Delta E(\omega)}
\label{Qdef}
\end{equation}

Where $E_{ave}$ is the average stored energy, $\omega$ the frequency, and $\Delta$E the energy lost per cycle. This differs slightly from definitions of $Q$ using $E_{max}$, the maximum stored energy per cycle, but following \citet{Bland1960} equation \ref{Qdef} is best compatible with the formal viscoelastic definition of $Q$ we discuss in section \ref{viscosection}, equation \ref{Qdefformal}.  Note that $Q$ will differ depending on the frequency for which it is measured,  as well as on the temperature and microdynamic properties of a material. 

For the whole Earth, a Quality factor of 12 to 34 \citep{Yoder1995, MurrayDermott2005} has been measured for the overall tidal process based on the expansion of the moon's orbit \citep{Dickey1994}. These values however include all damping effects of the lunar tide on the Earth, including the sloshing of Earth's oceans, and varied continental configurations. \citet{GoldreichSoter1966} estimate $Q$ for Mercury, Venus, and the Moon at $Q_{merc}$$\leq$190, $Q_v$$\leq$17, and 10$\leq$$Q_m$$\leq$150 respectively. \citet{Lainey2007} recently estimate $Q_{mars}$=79.91 ($\pm$0.69) from the motion of Phobos. Data summarized in \citet{KaratoSpetzler1990} suggest Earth's lower mantle $Q$ at tidal periods from 1-10 days is in the range 50-200. Widescale partial melting may also drive ocean-free 1$M_E$ $Q$'s closer to 10. For an evolved dry Earth-mass planet, we adopt a solid body baseline of $Q$=50. Extending such results to $Q$=10 or $Q$=100 is straightforward given $Q$ is linear in equation \ref{SnMEdot}.

The Love number $k$, is in essence a compliance term, characterizing the deformation response of a planet to stress.  As tides excite the $l$=2 prolate harmonic, we are concerned only with the $k_2$ term. A $k_2$ of 0 represents a perfectly rigid body, and 3/2 a perfect fluid. Earth has a value of 0.299 \citep{Yoder1995}. Through self-gravity, $k_2$ depends strongly on an object's size, with Mars at 0.14, and Jupiter still far from the fluid limit at $k_2$=0.38 \citep{Gavrilov1977}. Small strength dominated moons such as Phobos and Amalthea have Love numbers near zero. Thus for most terrestrial planets there is less than a factor of 2 uncertainty in $k_2$. 

As we proceed it is important to keep in mind errors incurred by our model choice. Equation 1 is based on linear tidal theory, and we are applying it to cases where dissipation is not small. Exoplanet stains in the region of interest for tidal heating range from 10$^{-8}$ to as high as 10$^{-4}$ or 10$^{-3}$ in extreme cases, depending on eccentricity, 
while nonlinearities may become important at strains greater than 10$^{-6}$. Corrections involving higher order terms and small angle approximations such as those discussed by \citet{EfroimskyWilliams2009} for tides in a primary, are more likely to become non-negligible for extreme tidal exoplanets. These corrections however may themselves be small compared to the greater error of not including planet-wide partial melting. 

Values of $k_2$=0.3 and $Q$=50 at 1$M_E$ allow us a simple baseline to highlight other features of the tidal heating parameter space. 

\subsection{Energy Balance}

To judge the geologic importance of tidal heating, we compare against other heat sources, specifically internal radiogenic heating and insolation from the host star. We estimate surface temperatures based on a blackbody assumption and a summation of primary energy sources: insolation, radionuclides, accretion (or gravitational) heat, and tides:

\begin{equation}
\dot{E}_{total} = \dot{E}_{insol}+ \dot{E}_{radio}+ \dot{E}_{grav}+ \dot{E}_{tidal}
\end {equation}

The blackbody surface temperature of such a model planet is found from the Stefan-Boltzmann Law 

\begin{equation}
\dot{E}_{total} = \dot{E}_{blackbody} = 4\pi R_{sec}^2 \sigma_{sb} \epsilon_r T_{surf}^4
\end {equation}

where $\sigma_{sb}$ is the Stefan-Boltzmann constant, and emissivity $\epsilon_r$ is close to unity (0.9 in the infrared). By considering planets as blackbodies we implicitly assume thin atmospheres and no significant greenhouse effects. 

We have tested models both with and without diurnal variations, since supertidal planets are likely to be spin locked to their host stars. Heliosynchronous spin locking may establish a strong boundary condition contrast for the mantle, of 1000K or more between dayside and nightside, if held stationary long enough. Such contrasts may alter convection plume patterns and induce true polar wander of vigorous elevated limb plumes back to the equator.

Equating $\dot{E}_{total}$ to the outgoing radiation is an assumption of steady state. The time it takes a given supertidal planet to reach equilibrium is discussed in section \ref{equilibria}, but is generally on the order of a few million years.

We scale insolation from the host star luminosity, $L_{star}$ in watts. Correcting for albedo $A$, the total insolation heat rate of a planet is:

\begin{equation}
\dot{E}_{insol} = (\pi R_{sec}^2)(1-A)\biggl(\frac{L_{star}}{4\pi a^2}\biggr) 
\end {equation}

This radiation its rapidly re-radiated at thermal wavelengths, but in so doing it adds to the outward blackbody flux, establishing a surface boundary condition for the planet's internal geotherm.

	Estimates for Earth's internal heat flux range from 30 to 46 TW \citep{Jaupart2007}. Measurements are made by sampling the geotherm in deep sea cores and equilibrated drill holes, as well as satellite IR photometry of the entire Earth. Much of the uncertainty comes from estimating the flux from volcanic point sources and through ocean crust hydrothermal circulation. About half of this heat is thought to come from radionuclides. The rest comes from several sources, including leftover accretion heating (gravitational potential heat trapped by burial), chemical phase change heat, primordial gravitational heat of the core differentiation event, continuing latent and gravitational heat release from the inner core crystallization, heat from the moon forming impact, and the pulse of heat provided by extinct short lived nuclides such as $^{26}$Al. In the deep Earth, tidal heating is negligible.
		
Although we are mostly concerned with $\sim$1$M_E$ exoplanets, for super-Earths we scale radiogenic and residual accretion heat by the total planet mass relative to 1$M_E$ (assuming a high silicate mass fraction). Using chondritic concentrations of $^{235}$U, $^{238}$U, $^{232}$Th, and $^{40}$K, we find radionuclide heating just under 9 times greater for an early Earth. Actual mantle nuclide concentrations are highly uncertain \citep{McDonoughSun1995}. However for tides to dominate an evolved planet they may need only exceed $\sim$40TW, but to dominate a younger planet just at the end of its migration phase, they may need to exceed $\sim$400TW. For radionuclide poor Ice Earths, smaller amounts of tidal heat will have greater consequence. We do not explore any range of core mass fractions or the possibility of radionuclide concentrations significantly different from our solar nebula. This in effect assumes a turbulent interstellar medium well mixed by random supernovae, and is generally supported by observations \citep{Elmegreen2004}. However, the time since the nearest supernovae can vary the concentration of $^{26}$Al and hence the initial pulse of heat to a planet. The occurrence or absence of an early giant impact might have a similar effect. 

\subsection {Fixed Q Results}

\begin{figure}[t]
\centering
\plotone{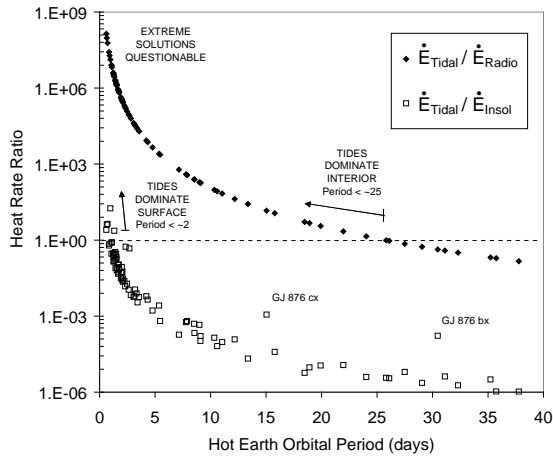}
\caption{Comparison of tidal heat to radiogenic and insolation heat for hypothetical Hot Earths in 2:1 resonance with known exoplanets. Body parameters including the modern radiogenic heat rate ($\sim$19TW) for Earth are used. Even given a conservatively strong forcing of $e$=0.1, tides only reach parity with insolation for extreme objects inside 2 day periods, where global-scale partial melting may invalidate the fixed Q assumption. 1$M_E$, $Q$=50, $k_2$=0.3, $A$=0.3.}
\label{ratios2}
\end{figure}

Figure \ref{ratios2} compares the ratios of tidal heat to insolation and radiogenic heat for hypothetical Hot Earths (designated by the suffix $x$) trapped in 2:1 resonances with known short period exoplanets as taken from the exoplanet.eu database of Jean Schneider. Scatter of the points is due to the varied luminosity of certain stars, with higher outliers being M dwarf hosts. A sufficient atmosphere is assumed to transport heat evenly to the nightside.  

Tidal heat dominates over radiogenic heat at periods under about 30 days, while it only dominates insolation below $\sim$2 days. This defines our range of interest for tides. When tides dominate radionuclides, they dominate the internal behavior of the planet. For super-Earths, large hosts, and uncommonly high eccentricities, the range of internal relevance shifts outwards, to typically no more than $\sim$80 day periods. When tides dominate over insolation, they can begin to be expressed in the observable temperature of the planet. 

These data show it is relatively easy for tides in short period objects to have a strong influence on internal activities, but it is very difficult for tides to matter for overall surface temperature. This is a consequence of the fact that to get strong tides in the first place, you have to be very close to a star, where $\dot{E}_{insol}$ $\sim$1$\times$10$^6$ to 1$\times$10$^8$ TW. At dimmer stars, tides can have greater surface significance. 

\begin{figure}[t]
\centering
\plotone{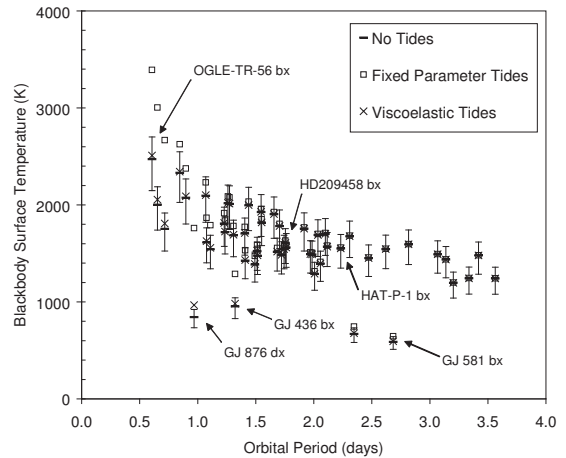}
\caption{Blackbody surface temperatures for hypothetical supertidal planets below 4 day orbital periods. Simulated data using the known catalog of exoplanets as 2:1 perturbers, with scaled stellar host masses and luminosities. Black marks with error bars: No Tides ($e$=0) with albedo uncertainty from $A$=[0.0, 0.3(nominal), 0.8]. M dwarf hosts form the low outliers. Squares: Fixed Parameter Tides results using 1$M_E$, $e$=0.1, $Q$=50, $k_2$=0.3. Crosses: Viscoelastic Maxwell rheology (more conservative estimates). Note tidal surface enhancements are generally lost in albedo uncertainty (even at this extreme $e$) except at dim stars and periods below 1 day.} 
\label{fig:BBsurftemps}
\end{figure}

Figure \ref{fig:BBsurftemps} shows tidal relevance with respect to surface temperature. Again, the point is that as an observable, the surface temperature of even extreme tidal objects is generally overwhelmingly dominated by insolation. Surface temperatures with and without tides are nearly equivalent except for the very nearest bodies around 1 day orbits, where the tidal signal reliably separates from the insolation signal. The plotted error bars represent an uncertainty range in albedo of 0 (dark ash filled skies) to 0.8 (bright solid clouds), with 0.3 (Earthlike) being the centermark. For anything but the most extreme cases or dimmest stars, uncertainty in albedo will prevent overall surface temperature from being used as a means to observationally identify planets as supertidal. Beyond 2 day orbits, the tidal contribution to blackbody temperature is negligible. Within 1 day periods, we argue that the tidal signal is erroneously high, due to the limitations of the fixed $Q$ model with no melting or viscoelasticity.
	
Even extensive volcanism induced by tides may have little impact for close planets of G class hosts, since the stars alone can maintain high enough temperatures to liquefy solid material. For brighter A and F stars tides should matter even less. The zero pressure melting point of many silicates lies between 1400 to 2100 K. Therefore from Figure \ref{fig:BBsurftemps}, insolation driven magma oceans are plausible, pending atmospheric effects, inside of 3 day G class orbits. The depth of such oceans will depend on the subsurface heat flux, ocean convection rate, composition, and pressure dependence of the solidus. For large planets such as Earth, pressure rises rapidly with depth and prevents melting even at very high temperatures. Tidal heat, though minor compared to insolation, may contribute to increasing otherwise shallow magma ocean depths. The behavior and tidal sloshing of these insolation established magma oceans, and how they contribute to the global $Q$ value is a subject of future work.
	
For M dwarf stars and nonluminous hosts, tides matter more. The reduction in tides due to a smaller host mass is easily outweighed by reduced insolation. It may be possible to observationally detect a few degrees tidal temperature enhancement at these dimmest stars. Note the data points labeled GJ581bx, GJ436bx and GJ876dx in Figure \ref{fig:BBsurftemps}. A broader axiom is that tides matter most in colder environments. A relatively low 20 TW tidal output at Io is strongly expressed due to the absence of strong radionuclide heat and comparatively weak insolation. Although this paper focuses on planets, we do expect tidal heating is of greatest consequence, creating new habitable regions, in cold exomoon systems analogous to Io, Europa, and Enceladus. 
   
For spin-locked objects, even if only a small fraction ($\sim$0.1-1\%) of insolation is transported to the nightside, large tidal contributions still appear negligible in most cases. If zero insolation is transported to the nightside, then opportunities do exist to detect the small tidal temperature elevations of 1-5 degrees that occur in orbits from 2-5 days, however this signal is still weak and such objects may be rare.   
   
Figure \ref{fig:BBsurftemps} suggests that to observationally confirm extreme tidal heating near stars one should look for spectroscopic evidence, such as magma hotspots or a chemical signature of widespread volcanism. Widespread devolatilization of a planet may produce extended $CO_2$ and sulfur compound absorption lines in observed spectra, L. Kaltenegger et al. (2010, in preparation). Light from hotspots and lava lakes may be polarized as on Io \citep{Veeder1994} and shifted to the near IR, aiding in detection. Clouds and haze may obscure hotspots, magma lakes, or even magma oceans, but if cloud breaks exist, then such observations are possible. In the meantime, eccentricity and resonances are the best indicators of tidal heating. Chemical and hotspot follow up observations can help in bounding global $Q$ values.
   
The most Io-like exoworlds will be those where insolation is too weak to melt the surface on its own, but tides are great enough to drive volcanism in excess of background rates. Consider that $\dot{E}_{tidal}/\dot{E}_{insol}$ and $\dot{E}_{tidal}/\dot{E}_{radio}$ are $\sim$0.08, $\sim$77 for Io, and $\sim$0.01, $\sim$10 for Europa. Thus dominance over insolation is unnecessary for Io-like supertidal properties. While moderate tides may not noticeably change a planet's average blackbody surface temperature, they can slow secular cooling and radically alter layer structure, plate tectonics, and the state of volcanism. Moderate tides, of a few hundred TW, are also less likely to break resonances or result in rapid circularization. 

The main point from this exercise that motivates our further work is the magnitude of tidal heating predicted by the fixed $Q$ and $k_2$ method. In Figure \ref{ratios2} the tidal heating of the shortest period hypothetical planets rises to over a million times the radiogenic rate: tens of millions of terawatts. The question we seek to explore is whether or not such extreme answers are possible. Can an Earthlike planet experience millions of terawatts in tidal output for any length of time? A number of physical effects conspire against this extreme answer: viscoelasticity, global-scale partial melting, convective feedback, and thermal-orbital coupling. Next in this paper we address viscoelasticity and melting.  

\section{Viscoelasticity}	
\label{viscosection}
	
Using equation \ref{SnMEdot} to calculate global tidal heat is useful for estimates, however it ignores the frequency dependence of a material's response to loading. $Q$ and $k_2$ are neither constant nor entirely independent parameters, \citep{Kaula1964, Zschau1978, Segatz1988}. In its most general form for a viscoelastic body, the ratio $k_2/Q$ is replaced by the imaginary part of the complex Love number $-Im(k_2)$, which characterizes the material's viscous phase lag. 
  
\begin{equation}
\label{HnSEdot}
\dot{E}_{tidal}=-Im(k_2)\frac{21}{2} 
\frac{G M_{pri}^2 R_{sec}^5 n e^2}{a^6}
\end {equation}

This formula still assumes a homogeneous body. A complete calculation of tides would consider variations by layers using a propagator matrix method \citep{Takeuchi1962} as well as the full three dimensional stress and strain tensors to compute tides as a function of latitude and longitude \citep{PealeCassen1978, Segatz1988}. However equation \ref{HnSEdot} is effective in seeking estimates and extrema of a globally averaged behavior. 

In general, $Im(k_2)$ will have a response peak at a material's characteristic natural frequency, allowing bodies to tidally resonate. To compute $Im(k_2)$ for any rheology, we begin from a definition of the Love number:

\begin {equation}
\label{lovedef}
k_2 = \frac{3}{2}\frac{1}{1+\tilde{\mu}}
\end {equation}

Where $\tilde{\mu}$ is known as a body's effective rigidity

\begin {equation}
\label{mutilde}
\tilde{\mu} = \frac{19\mu}{2\rho g R_{sec}}
\end {equation}

and $\mu$ is a material rigidity. Effective rigidity is a non-dimensional ratio that compares elastic forces in the numerator to gravitational forces in the denominator. For $\tilde{\mu} >> 1$ the body is strength dominated, while for $\tilde{\mu} << 1$ it behaves like a gravitating fluid. Often the substitution 

\begin {equation}
\beta = \rho g R_{sec} 
\end {equation}

is made to emphasize this point by forming a gravitational stiffness analogous to the material modulus $\mu$. Since tides excite a planet in shear (orthogonal tension and compression), for $\left|\mu\right|$ we use the Shear modulus $G$ (not the Bulk or Young's modulus $K$ or $E$). Typical values of $G$ are 90 GPa for (undamaged) rocky material, and 4 GPa for icy material \citep{Goldsby2001}. In comparison, the effective gravitational rigidity is $\beta$=250 GPa for Earth and 8 GPa for Io. 

A material may behave as any combination of springs and dampers that model viscous creep and elastic rebound. Mathematically, a spring damper model is represented by forgoing a scalar modulus $\mu$ for a complex value $M^{\ast}$, whose real and imaginary parts define the energy storage and energy loss aspects of the system. These components $M_1$ and $M_2$ are computed from a model's constitutive equation. Substituting the complex form of the stiffness 

\begin {equation}
\mu(\omega) = M^{\ast}(\omega) = \frac{\sigma(\omega)}{\epsilon(\omega)} = M_1(\omega) + \imath M_2(\omega) 
\end {equation}

into equation \ref{mutilde}, then substituting this complex from of $\tilde{\mu}$ into equation \ref{lovedef}, we derive the complex form of $k_2$ and extract the imaginary component $Im(k_2)$ for use in equation \ref{HnSEdot}. 
	    
\subsection{Anelastic Models}

\begin{table*}
\begin{center}

\fontsize{8pt}{10pt}\selectfont

\caption{Viscoelastic Formulae}
\label{eqntable}
\begin{tabular}{|c|c|c|c|}
\hline
&Maxwell&	SAS&	Burgers$^\ast$\\
\hline
$\tau$ \T \B &
$\frac{\eta}{M}$ \T \B &
$\eta \delta\!J$ \T \B &
$\tau_A = \eta_A \delta\!J$,   $\tau_B = \frac{\eta_B}{M_B}$ \T \B \\
\hline
$M_1$ \T \B &
$\frac{M\eta^2\omega^2}{M^2+\eta^2\omega^2}$ \T \B & 
$\frac{1}{J_R} + \frac{(\eta \omega)^2 \delta\!J^3 J_u}{J_R^3+ J_R(\eta \omega J_u \delta\!J)^2}$ \T \B &
$\frac{\omega^2(C_1-\eta_A \delta\!J C_2)}{C_2^2 + \omega^2 C_1^2}$ \T \B \\
\hline
$M_2$ \T \B &
$\frac{M^2\eta\omega}{M^2+\eta^2\omega^2}$ \T \B & 
$\frac{\eta \omega \delta\!J^2}{J_R^2+(\eta\omega J_u \delta\!J)^2}$ \T \B &
$\frac{\omega(C_2+\eta_A \delta\!J \omega^2 C_1)}{C_2^2 + \omega^2 C_1^2}$ \T \B \\
\hline
$-Im(k_2)$ \T \B &
$\frac{57 \eta \omega}{4\beta(1+[(1+(19M/2\beta))^2\eta^2\omega^2/M^2])}$ \T \B & 
$\frac{57 \delta\!J^2\beta\eta\omega}{(19+2\beta J_R)^2+\delta\!J^2\eta^2\omega^2(19+2\beta J_u)^2}	$ \T \B &
$\frac{57 \beta \omega (C_2 +\delta\!J \omega^2 C_1 \eta_A)}{361\omega^2+76\beta \omega^2C_1+4\beta^2\omega^2C_1^2+(2\beta C_2-19\delta\!J\omega^2\eta_A)^2 }	$ \T \B \\
\hline
$Q$&
$\frac{\eta \omega}{M}$ \T \B &
$\frac{J_u + \delta\!J + J_u(\omega \eta \delta\!J)^2}{\eta \omega \delta\!J^2}$ \T \B &
$\frac{\omega (C_1 - \eta_A \delta\!J C_2)}{C_2 + \eta_A \delta\!J \omega^2 C_1}$ \T \B \\
\hline
&
\multicolumn{2}{c|}{
$ ^{\ast}$  
$C_1 = \delta\!J + \delta\!J \frac{\eta_A}{\eta_B} + \frac{1}{M_B} \T \B $  
}&
\multicolumn{1}{c|}{
$ ^{\ast}$
$C_2 = \frac{1}{\eta_B} - \frac{\eta_A \delta\!J}{M_B} \omega^2 \T \B $ 
}\\
\hline
\end{tabular}
\end{center}
\tablecomments{For the SAS model $J_u$ represents the initial elastic compliance, $\delta\!J$ is the additional later compliance due to creep, and $J_R = J_u + \delta\!J$ is the relaxed compliance following creep. Burgers formulas are expressed in terms of the two lumped terms $C_1$ (which is an effective compliance) and $C_2$.}
\end{table*}

\begin{figure}[t]
\centering
\plotone{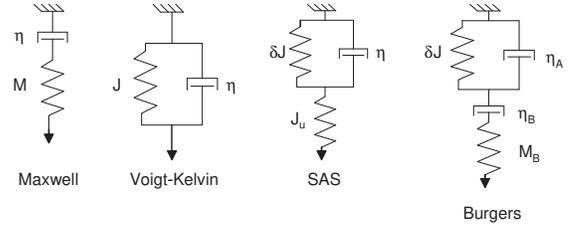}
\caption{The four different anelastic models discussed in this paper, and their corresponding notation.}
\label{fig:anelasticmodels}
\end{figure}

  The four basic models and parameters for the rheologies considered for this paper are shown in Figure \ref{fig:anelasticmodels}. The Maxwell body, commonly used because of its simplicity, considers mantle rock as a spring-dashpot series, with an instantaneous elastic response, followed by viscous yielding. It is ultimately fluid. 
  
The Maxwell model is useful but incomplete \citep{Ojakangas1989}. As discussed by \citet{Zener1941} and observed in the laboratory \citep{Post1977, SmithCarpenter1987, Cooper2002, Jackson2000}, real polycrystalline materials exhibit a wider range of relaxation mechanisms. 
	
A parallel spring-dashpot pair is known as the Voigt-Kelvin model. Here viscous relaxation is ultimately limited by the spring. While the Voigt-Kelvin model is instructive as a subcomponent of other models and has been applied to lunar tides \citep{Vanarsdale1981}, we ultimately find it poorly suited to short period cases.
	  	
A three parameter model, known as either the Standard Anelastic Solid, (SAS) or Standard Linear Solid, has features of both the Maxwell and Voigt-Kelvin primitives: instantaneous elastic response, followed by strain limited relaxation. It will not take a permanent set. All deformation is recovered when a load is removed. Either of the two ways to arrange two springs and one damper in a series-parallel combination are mathematically equivalent \citep{NowickBerry1972}.
	
A four parameter model, or Burgers body, allows the modeling of transient molecular creep behavior in minerals. It can exhibit transient creep, recovery, and take on a permanent set, modeling a broad range of materials. The Burgers or SAS models may both be reduced to the Maxwell or Voigt-Kelvin models through appropriate selection of parameters. 

The Burgers body is useful in modeling the phenomenon of grain-boundary slip. The Maxwell element within the Burgers body represents classical diffusion creep, where non-recoverable creep motion occurs through void migration inside of grains. Grain boundary slip occurs on a shorter relaxation timescale and is recoverable, as represented by the Voigt-Kelvin element. Postglacial rebound studies in particular have suggested that the Burgers body is a more appropriate model of the Earth than a Maxwell body \citep{Sabadini1987, FaulJackson2005}. As we apply them to exoplanet tidal models, both the SAS and Burgers models reveal susceptibility modes not found in a Maxwell approach. We note that as a method of extrapolation, spring-dashpot models have the advantage of a clear relationship to underlying defect microdynamics, but the disadvantage of historically poor correlation with large complex inhomogeneous systems such as Earth's mantle.

We follow the notation of \citet{NowickBerry1972} where $J$'s denote compliances in $Pa^{-1}$, and $M$'s denote stiffnesses in $Pa$.  Series springs are mainly expressed by stiffnesses and parallel springs by compliances. For the SAS model $J_u$ is the instantaneous compliance to an applied load, and $\delta\!J$ is the additional compliance during creep, known as either the creep defect or compliance defect. 

Using the geometry and parameters of each model from Figure \ref{fig:anelasticmodels}, then solving for the overall stress $\sigma$ and strain $\epsilon$ leads to the following constitutive relations:

\vspace{2 mm}

Maxwell
\begin{equation}
M\sigma + \eta \dot{\sigma} = M\eta \dot{\epsilon}
\end{equation}

Voigt-Kelvin
\begin{equation}
J\sigma = \epsilon + \eta J\dot{\epsilon}
\end{equation}

SAS
\begin{equation}
(J_u + \delta\!J)\sigma + \eta \delta\!J J_u \dot{\sigma} = \epsilon + \eta \delta\!J \dot{\epsilon}
\end{equation}

Burgers
\begin{equation}
\frac{\sigma}{\eta_B} + \left(\delta\!J + \delta\!J \frac{\eta_A}{\eta_B} 
+ \frac{1}{M_B}\right)\dot{\sigma} + \frac{\eta_A \delta\!J}{M_B}\ddot{\sigma}
= \dot{\epsilon} + \eta_A \delta\!J \ddot{\epsilon}
\end{equation}

While a tidal distortion of a planet is best expressed as an applied strain, solutions were also investigated for applied stress problems. The main difference between these methods is that the Maxwell and Voigt-Kelvin models switch behaviors. In a Maxwell model there is nothing to prevent infinite viscous work under an applied stress, as no spring limits the dashpot's travel. Under an applied strain, work is inherently finite. The reverse is true in a Voigt-Kelvin model, where rapid forced strain cycles can drive the dashpot to approach infinite work. Mainly for this reason we find the Voigt-Kelvin model unsuited to tidal cases, and the most insight is found from comparing only the Maxwell, SAS and Burgers models under a cyclic applied strain.  

The generalized equations for cyclic applied strain forcing are:

\begin{equation}
\label{appstrain1}
\sigma(\omega) = (\sigma_1 + i \sigma_2) e^{i \omega t}
\end{equation}

\begin{equation}
\label{appstrain2}
\epsilon(\omega) = \epsilon_o e^{i \omega t}
\end{equation}

Taking the first two time derivatives of equations \ref{appstrain1} and \ref{appstrain2}, then substituting these derivatives into the constitutive equations, leads to the complex stiffnesses $M_1$ and $M_2$ as they are reported in Table \ref{eqntable}.   

Table \ref{eqntable} also reports the quality factor, relaxation timescale $\tau$, and $Im(k_2)$ formulae of each model. The Quality factor $Q(\eta,\omega)$ is found as the ratio of the compliances \citep{OconnellBudiansky1978}, and is the same value under either applied stress or strain.

\begin{equation}
Q(\eta,\omega) = \frac{J_1(\eta,\omega)}{J_2(\eta,\omega)}= \frac{M_1(\eta,\omega)}{M_2(\eta,\omega)}
\label{Qdefformal}
\end{equation}

Work per cycle per unit volume for a generic object can be found from the integral of stress times strain. In a sinusoidal uniaxial applied strain case:

\begin{equation}
W(\eta(T),\omega) = \pi M_2(\eta(T),\omega)\epsilon_o^2
\end{equation}

This formula is nearly equivalent to equation \ref{HnSEdot} once tidal strain is computed from the orbit, however it omits key ingredients, specifically: self-gravity, a shape factor of 2/5 for a sphere, and a factor of 7/3 for the combination of what are termed the radial and librational tides (arising from spin synchronization in an elliptical orbit). Self-gravity in effect acts like an additional spring in all of the diagrams in Figure \ref{fig:anelasticmodels}.  When a bulge is raised on a planet's surface, both gravity and material rigidity act to pull the bulge back down. To properly include self gravity, we use equation \ref{HnSEdot} and express complex compliances in terms of $Im(k_2)$ following equation \ref{lovedef} and the method outlined at the head of this section. These results for $Im(k_2)$ are given in Table \ref{eqntable}.

Using equation \ref{HnSEdot}, $W(\eta(T),\omega) = \dot{E}_{tidal}$ is entered into a global energy balance along with radionuclides, convection, melting, and a viscosity model, then propagated in time to observe the temperature behavior of the planet.  

\subsection{Viscosity Model}

For a viscosity model we use an Arrenhius relation

\begin{equation}
\label{arrhenius}
\eta(T) = \eta_o e^{\frac{E^{\ast}} {RT} }
\end{equation}

where $R$ is the universal gas constant, $E^{\ast}$ an activation energy, and $\eta_o$ a defining viscosity. 

From isostatic rebound studies of Canada and Scandinavia \citep{MitrovicaForte2004}, Earth's viscosity is of the order $\sim4\times10^{20}$ to $\sim1\times10^{23}$ Pa$\cdot$s from the upper to lower mantle respectively. Results can be highly sensitive to this parameterization. We choose $\eta_o$ to match a selected setpoint $\eta_{set}$, either $1\times10^{22}$ Pa$\cdot$s at 1000K (a weaker or wetter case), or $1\times10^{24}$ Pa$\cdot$s at 1000K (a stronger or more devolatilized case). Higher $\eta_{set}$ choices in general attenuate the global response. Liquid phase viscosity is modeled with magmas near 1000 Pa$\cdot$s at 2000K \citep{McBirney1984}.

The energy barrier for some combination of creating and moving lattice defects in creep is characterized by $E^{\ast}$. Values relevant for dislocation creep at the stresses and pressures of Earth's mantle are from 300 to 400 kJmol$^{-1}$ \citep{Ashby1977, FischerSpohn1990}. At very low stresses the rheology may switch to diffusion creep, either Coble or Nabarro-Herring based on the temperature \citep{FrostAshby1982}, however we assume the stress state for supertidal planets remains high enough (1 GPa or more) to neglect these creep mechanism transitions. In addition we neglect pressure variation of $E^{\ast}$ and changes due to varied concentrations of dissolved volatiles. 

Baseline parameters are summarized in Table \ref{paramtable}. Given $\eta(T)$ it is possible to replace the constant terms $\eta$ in Table \ref{eqntable} and obtain tidal work as $W(\eta(T), \omega)$. 

While not modeled in detail here, ice viscosities range from $1\times10^{11}$ to $1\times10^{14}$ Pa$\cdot$s \citep{Poirier1981}. With a 4 GPa rigidity, this produces Maxwell times from a few minutes to a year. Thus Ice Earths are strong candidates for a peak viscoelastic tidal response. In addition to warming icy moons outside the snowline, tides may play a unique role in the degradation of ice mantles on large migrated inner solar system objects, as a way to emplace heat beneath thick high albedo cloud decks. 

Earth's bulk resonant Maxwell timescale $\tau$ (see Table \ref{eqntable}) is a few thousand years, fairly unresponsive to a tidal period of a few days, and with an attenuating $Im(k_2)$ factor of $\sim9\times10^{-6}$. Io at 1.7 days is suggested to be nearer its Maxwell response peak, meaning a mantle viscosity $\sim1\times10^{16}$ Pa$\cdot$s, perhaps due to temperature, partial melting, and composition. Earthlike bodies may require a heat pulse trigger, or pre-softening, to reach such a tidally responsive Io-like state. Impacts, impact induced librations, and impact induced obliquity tides have been suggested as triggers for tidal heating in Saturnian satellites \citep{Castillo2006a, Castillo2006b}. Post-migration planets may also experience tidal heat while still hot from accretion and $^{26}$Al. Once well coupled, tides have the potential to keep mantles warm thereafter. Alternatively, such triggers may not be needed to strongly couple tides in the SAS and Burgers models, where response peaks occur at lower temperatures due to the term $\delta\!J$.

\subsection{Melting Model}

A description of silicate melting allows us to resolve both the rapid increase in convective vigor and the decoupling of tides that simultaneously occur when viscosity and shear modulus decrease. Parametric models of melting for Io are presented by \citet{Moore2003b} and \citet{FischerSpohn1990} based on laboratory experiments by \citet{Berckhemer1982}. These models variously represent the essential feature of a breakdown temperature: at some point in the partial melting (or crystallization) process, a material switches from being best described as a solid matrix with fluid pores, to a fluid bath with isolated floating crystals grains. When grains loose contact with one another, the material looses shear strength and switches to the viscous properties of the fluid. 

We follow the \citet{Moore2003b} model melting values of $T_{sol}$ = 1600K, $T_{liq}$ = 2000K, and $T_{brkdwn}$ = 1760K (40\% melt), 1800K (50\%melt), or 1840K (60\% melt), along with a mantle specific heat $C_p$ = 1260 J/kg-K, and a heat of fusion $H_f$ = 500,000 J/kg. Shear modulus $G$ is constant up to the solidus then begins to drop according to an empirical Arrhenius law 

	\begin{equation}
			 G =  G_o e^{(\frac{40000}{T}-25)}
	\end{equation}

with an activation temperature 40000K. Upon breakdown, the shear modulus drops to zero, and the viscosity immediately becomes that of the liquid. Between the solidus and the breakdown temperature, viscosity drops off by a power law
	
	\begin{equation}
			 \eta =  \eta_{solid} e^{-40 \chi}
	\end{equation}
	
where $\chi$ is the percent melt fraction and $\eta_{solid}$ is the viscosity at the solidus temperature. Moore also presents a stiffer material where the empirical coefficient -40 is replaced by -10.
	
The primary shortcoming of our melting approach is applying it to the planet as a whole. This model assumes a uniform adiabatic mantle temperature, a single solidus temperature, uniform global onset melting, and ignores compositional, pressure, eutectic, and hydration effects on the solidus. This model does allow us to capture the gross behavior of tidal shutdown at some small degree of partial melt, without telling us how the melt distribution affects the exact tidal result. In section \ref{discussion} we discuss the role of local partial melting as well as the role of heat transport by advective magma percolation.

\begin{table}
\begin{center}
\caption{Baseline Material Parameters} 
\label{paramtable}
\begin{tabular}{lll}
\hline  
Series Shear Modulus&       $M$, $M_B$ =&               $5\times10^{10}$ Pa\cr 
Creep Compliance&           $\delta\!J$ =&              $4\times10^{-12}$ Pa$^{-1}$\cr 
Defining Viscosity&         $\eta_{set}$ =&             $1\times10^{22}$ Pa$\cdot$s\cr 
Burgers Parallel Viscosity& $\eta_{Bset}$ =&            $2\times10^{20}$ Pa$\cdot$s\cr   
Activation Energy&          $E^{\ast}$, $E^{\ast}_B$ =& $300$ kJmol$^{-1}$\cr  
Solidus Temperature&        $T_{sol}$ =&                $1600$ K\cr
Liquidus Temperature&       $T_{liq}$ =&                $2000$ K\cr 
Breakdown Temperature&      $T_{brkdwn}$ =&             $1800$ K\cr 
\hline
\end{tabular}
\end{center}
\tablecomments{Used for reported heat rates unless otherwise specified.}
\end{table}

\subsection{Viscoelastic Results}
\label{viscoresults}

\begin{figure*}[t]  
\centering
\plotone{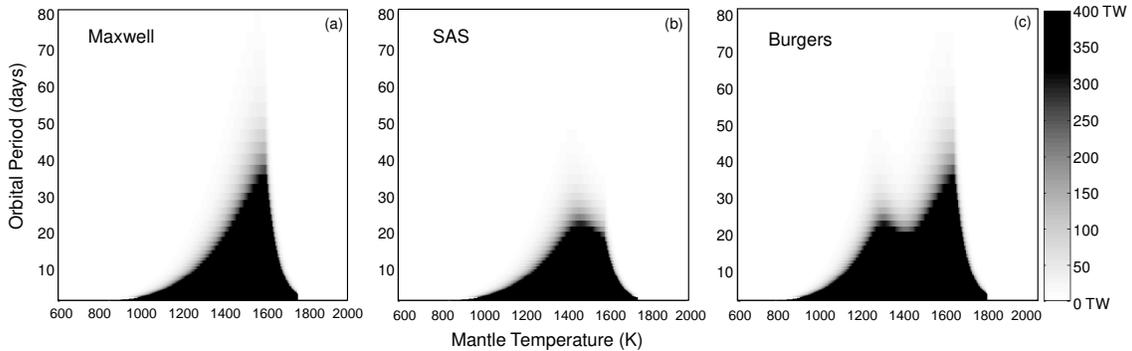}
\caption{Maps showing the magnitude of tidal heating (in TW) as a function of mantle temperature and orbital period for homogeneous planets of 1$M_E$ and $e$=0.1 around a solar mass host. Viscosity setpoint is $1\times10^{22}$ Pa$\cdot$s at 1000K. The solidus and liquidus are set at 1600 and 2000K respectively. Tidal heat diminishes rapidly above the solidus and is absent above the 1800K breakdown temperature. Broad regions of negligible and extreme heating are separated by narrow regions of geologically moderate heat. Gradients are clipped at 400 TW.} 
\label{tempperiodmap}
\end{figure*}

Figure \ref{tempperiodmap} compares the tidal work response $W(T,\omega)$ of the Maxwell, SAS and Burgers models in maps of temperature-period space. The figure can be read in horizontal slices, representing the temperature evolution of a planet in a fixed orbit. Or it may be read vertically, representing a planet migrating to shorter periods, encountering higher tidal heating as it moves, then shifting to higher temperatures in response. The peaks seen in each map are a kind of planetary resonance, akin to the Debye peak in frequency space. In this case they occur as changing mantle temperature in effect tunes the planet's response frequency to match its orbital period. Since tuning is linear in frequency but exponential in viscosity, the peaks appear at roughly constant temperatures.

Tidal output spans a wide range of magnitudes due to the large exponents in equation \ref{HnSEdot}, however Figure \ref{tempperiodmap} is linear and clipped to a maximum of 400 TW. This emphasizes the geologic impact of tides, as values much below a $\sim$40 TW radiogenic background are of little concern, while extreme values above a few hundred TW will move the system rapidly in temperature towards global partial melting. Note that regions of both negligible (white) and extreme tidal heating (black) are quite broad, while the geologically moderate range from 10 TW to 400 TW is comparatively narrow, suggesting that supertidal planets only rarely result in enduring Earthlike features such as plate tectonics. In section \ref{equilibria} we show how moderate cases may still be favored by equilibrium considerations.  

This figure is drawn for a high $e$=0.1, where significant heating extends out to 40-70 day orbits. Higher planet and host masses will extend the viscoelastic range of tidal relevance, while lower eccentricities pull the range of tidal relevance back within 30 days, as in section \ref{fixedQ}. 

Unlike the fixed $Q$ approach, Figure \ref{tempperiodmap} also shows that declaring a tidal output for a given planet knowing its period and eccentricity is not possible without knowing its internal mantle state. The range of uncertainly in heat due to temperature often spans from negligible to extreme. 

On one hand the Maxwell result is similar enough to the SAS and Burgers results that it may be sufficient to broadly categorize a planet. On the other hand, results at a given phase space location can differ by several orders of magnitude between models. In Figure \ref{tempperiodmap} full rounding of the SAS response peak is expressed since it is shifted fully to the cool side of $T_{sol}$. This is due to the SAS model's use of the creep compliance term $\delta\!J$, which is set to 20\% below the elastic compliance $J$, following data in \citet{SmithCarpenter1987}. Still, the run-up alone to the Maxwell peak here extends tidal influence out further than the SAS model, again due to using $J$ vs. $\delta\!J$. Placing a Maxwell peak within our interest range can be parametrically difficult without significantly lowering $\eta_o$.
 
The Burgers map is significantly different, with two response peaks in the temperature axis. The two peaks are well separated with $\eta_B$ and $\eta_A$ set to be a factor 50 apart (e.g. grain boundaries relax 50$\times$ faster than voids diffuse). This factor for a bulk planet is uncertain, with laboratory and model values in the range 2 to 100 \citep{Sabadini1987, Peltier1985}. Smaller values place the response peaks nearer together. Following \citet{Post1977} we let $E^{\ast}_A$=$E^{\ast}_B$, however small changes can significantly alter the peak spacing. 

More advanced models than the Burgers body exist, including ones with discrete Voigt-Kelvin subunits in series, and the Andrade model, where viscosity and shear are treated as continuum functions in frequency. Response peaks in real materials can be hard to isolate \citep{Cooper2002} and real planets will have a range of compositions and grain sizes, blurring an ideal response function. In such cases the importance of the Burgers model is less the gap between the peaks and more the overall response broadening. 

Existing dissipation measurements from Earth's Chandler wobble, tides and free oscillations, summarized in \citet{KaratoSpetzler1990} show $Q$ values which change gradually with frequency, and are modeled well by empirical power laws of the form $Q \sim \omega^{\alpha_Q}$ ($\alpha_Q$ = 0.1 to 0.4) \citep{Karato2007, EfroimskyLainey2007}. Thus, while a given Maxwell parameterization may match observations well at a 1 year period, it would then typically underpredict dissipation at 1-10 days by more than an order of magnitude. This is due to the linearity of $\omega$ in the Maxwell approach, leaving it far more sensitive to frequency changes than $\omega^{0.1}$ or $\omega^{0.4}$. While similar discrepancies may be expected in other mechanism based models, the response broadening in the Burgers model does help in alleviating $\omega$ sensitivity. In general, response broadening may be achieved by adding Maxwell elements to a model to represent new dissipation modes. For example \citet{OconnellBudiansky1977} show that fluid flow between cracks and pore spaces of varied size can lead to roughly constant $Q$ values over several decades in frequency. An Andrade model, either based directly on whole Earth data, or based on a continuum model of springs and dashpots to model the broad range of compositions in the Earth, may overcome many of these issues.

For this reason, the fixed $Q$ approach of section \ref{fixedQ} is valuable in worlds that are particularly Earthlike. The value of derived Maxwell, SAS, and Burgers models comes in extrapolating to new bodies, such as super-Earths, Hot Earths, Ice Earths, and Earths with varied compositions, varied partial melt fractions, extensive devolatilization, or static lids.
 
\begin{figure*}[t]
\centering
\plotone{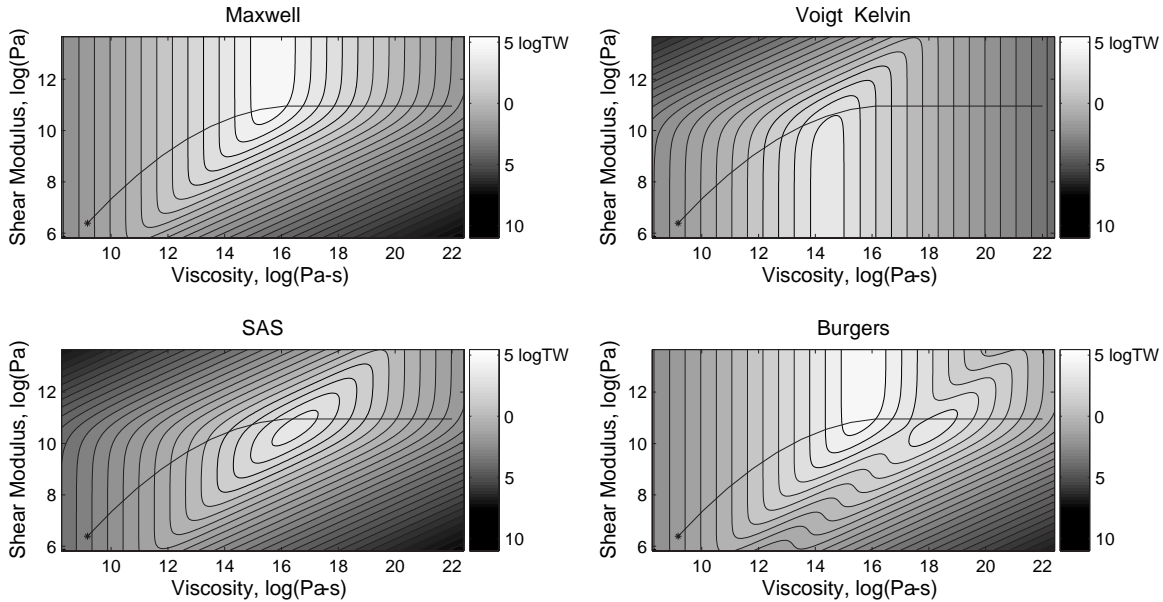}
\caption{Maps showing the magnitude of tidal heating as a function of shear modulus and viscosity for homogeneous planets of 1$M_E$, $e$=0.1, and a period of 10 days. Melting trajectories are overlain, and breakdown is marked with a dot. Changing the orbital frequency shifts the underlying map relative to given fixed melting path. The generally unsuitable Voigt-Kelvin model is included for comparison. Note how in the Burgers model a cooling planet can cross a radically different history of islands than in the Maxwell or SAS models. } 
\label{shearviscmap}
\end{figure*}

Many variations of Figure \ref{tempperiodmap} exist. In particular, altering the defining viscosity from $1\times10^{22}$ to $1\times10^{24}$ Pa$\cdot$s at 1000K shifts the response peaks to above the solidus, where they appear sharply compressed along the abscissa prior a falloff at $T_{brkdwn}$. Figure \ref{shearviscmap} helps explain this behavior.  

Figure \ref{shearviscmap} compares the models as maps in shear-viscosity phase space. Overlain on these maps are trajectories as a planet evolves in temperature. The upper branch represents the planet prior to melting, when shear modulus is constant. Evolving to the left is akin to heating with decreasing viscosity. Melting begins at the elbow, and the lower partial melt branch represents the fall of shear modulus and viscosity together. The slope of this branch is only moderately constrained by laboratory experiments. 

Beyond the solidus, a given $W(T, \omega)$ function will either show a sharply compressed response peak, or a dropoff with no peak. These maps help explain why. If the downward trajectory crosses a tidal response ridge or island, a sharp post-solidus peak will occur. If the downward trajectory slips off the side of an island, or passes between the saddle of the two Burgers islands, no post-solidus tidal peak occurs. 

Both Figure \ref{tempperiodmap} and \ref{shearviscmap} support the notion that a broad range of models and parameters result in significant phase space volumes where tidal heating is extreme, coexisting with broad negligible regions. We have explored variations of the Burgers parameters $\eta_A$ and $\eta_B$ to verify they move and stretch the dual features of the Burgers plots, but preserve the function's basic form. Variations in the creep activation energy $E^{\ast}$ (A and B for Burgers) has a strong impact on final magnitudes, but again smoothly deforms the functions. 

Because of the overall uncertainty in both parameters and models, we find it misleading to report a specific viscoelastic tidal output for a given planet without detailed analysis. Of more importance is the broad range of outputs possible. However, due to feedback with convection, a planet is not equally likely to reside at all possible temperatures, and instead will seek out specific equilibria.

\section{Tidal Equilibria}
\label{equilibria}

Following the method of \citet{Moore2003a, Moore2003b} for Io, we can model the thermal behavior of a supertidal body in time by simultaneously plotting tidal work as a function of temperature $W(T)$ atop convective output $\dot{E}_{conv}(T)$. Equilibria points occur whenever tidal input equals convective output. 

We use the method of parameterized convection \citep{OConnellHager1980} to iteratively solve for cooling as a function of mantle temperature. Convective vigor determines the thickness of a conducting boundary layer $\delta$ by the formula: 

	\begin{equation}
  \delta = \frac{d}{2 a_2} \left(\frac{Ra}{Ra_c}\right)^{-\frac{1}{4}}
  \end{equation}

$a_2$: Flow geometry constant $\sim$1 

$d$: Mantle thickness $\sim$3000km

$Ra_c$: Critical Rayleigh Number (free top)$\sim$1100
\\
 
where the exponent $-1/4$ is true for bodies with internal heat generation ($-1/3$ for bodies without). The Rayleigh number is found from the material properties of the uniform interior and heat flow through the top $q_{BL}$: 
	
	\begin{equation}
			Ra = \frac{\alpha \ g  \ \rho \ d^4 \ q_{BL}}    {\eta(T) \ \kappa \ k_{therm}}	    
  \end{equation}

$g$: Gravitational acceleration $\sim$9.81 m/s$^2$

$\alpha$: Thermal expansivity, $\sim$1x10$^{-4}$\citep{Kaula1968}

$\rho$: Average Density, $\sim$5000 kg/m$^3$

$\kappa$: Thermal diffusivity = $k_{therm}/\rho C_p$

$k_{therm}$: Thermal conductivity, $\sim$2 W/mK
\\	

Because of the viscous term $\eta(T)$, the Rayleigh Number responds rapidly to temperature. We neglect the weaker temperature dependencies in $\rho$ and $\alpha$. Conduction through the top boundary is computed based on an average thermal conductivity $k_{therm}$ and the thermal gradient.

\begin{equation}
		q_{BL} = k_{therm} \left(\frac{T_{mantle}-T_{surf}}{\delta(T)}	\right)		
\end{equation}
	
An initial guess is made for boundary layer thickness, then the model is iteratively solved for $q_{BL}(T)$. We assume the planet is in equilibrium with its star and has a thin atmosphere, allowing all heat that moves through the boundary layer to escape into space. 

\begin{figure*}[t]
\centering
\plotone{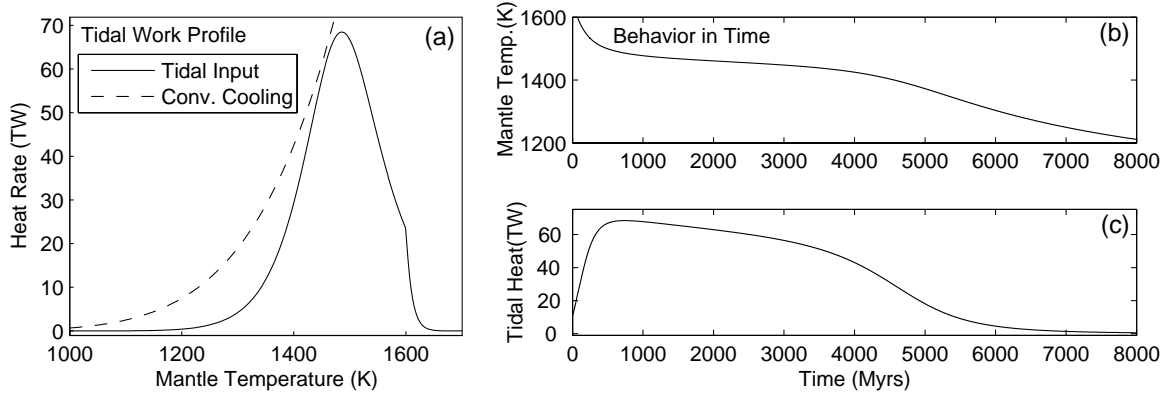}
\caption{Behavior Example 1. SAS Model. Nearly halted secular cooling. Panel (a): Tidal work profile (solid line) vs. convective output profile (dashed line) ($e$=0.044, 1$M_E$, 20 day period, material properties from Table \ref{paramtable}, hot start, negligible radionuclides, 0.5$M_{Sol}$ primary). Panels (b) and (c): Bulk mantle temperature and tidal heat evolution in time. Eccentricity is set for peak tidal heating to almost match convective cooling. Note the long timescale and long plateau as the system slowly evolves over the response peak. This forms one limit of slow secular cooling. Higher tidal forcing leads to tidal-convective stabilization and a trivial equilibrated time history. Lower tidal forcing leads to a weaker single episode of heating. Cold initial conditions will cool without peak crossing.}
\label{nearlyhalted}
\end{figure*}

\begin{figure*}
\centering
\plotone{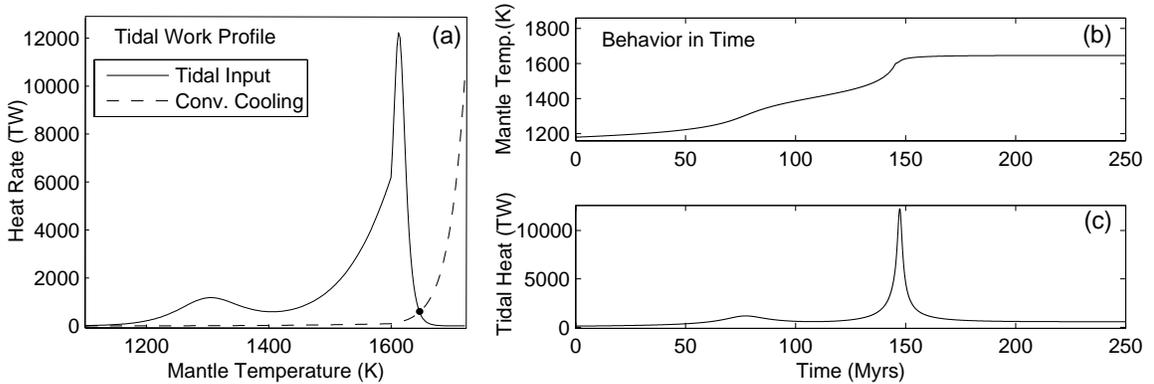}
\caption{Behavior Example 2. Burgers Model. Panel (a): Tidal and convective profiles. Hot stable equilibrium marked with solid circle. ($e$=0.05, $1M_E$, 12 day period, material properties from Table \ref{paramtable}, cool start, 1$M_{Sol}$ primary). Panels (b) and (c): Dual peak crossing: this strong forcing case heats the planet through both Burgers body resonance peaks, such that two unequal sudden warming episodes occur in time. Note the short timescale. Peak heating is $\sim$12200 TW while equilibrium occurs at $\sim$580TW and 1646K (Bulk $\chi$=11.5\%). Hot start planets would cool to the hot stable equilibrium without encountering the peaks of this scenario.}
\label{burgersbehavior}
\end{figure*}

Knowing the shape of $W(T)$ (plus a small radionuclide background) and $\dot{E}_{conv}(T)$ essentially characterizes the behavior range of a supertidal planet in time. The crossing points of the two functions represent equilibria points. These may either be stable or unstable based on the relative values of $dW/dT$ and $d\dot{E}_{conv}/dT$. Figure \ref{nearlyhalted} shows an example for mild SAS tides. Figure \ref{burgersbehavior} shows a stronger example for a Burgers body.  Young planets subject to tidal heating early may evolve into this system from high temperatures, while planets that significantly cooled prior to onset tidal forcing approach equilibria from the left. 

When tidal forcing is strong a hot stable equilibrium point typically exists near the onset of melting, where viscosity falls off sharply. Tides and convection cross here, with tides shutting down just as convection sharply ramps up. The location of this equilibrium is robust in cases when it occurs prior to onset melting. In cases where it occurs after melting onset, due to our homogeneous mantle assumption, only the existence of the equilibrium is robust. The predicted degree of partial melt $\chi$ is at best suggestive of how far a real mantle will melt. Once at a stable equilibrium, a planet will remain there at constant bulk temperature as long as the orbit allows.  

If tides are weak, $W(T)$ may never intersect $\dot{E}_{conv}(T)$, meaning no equilibria exist, and secular convective cooling will proceed uninterrupted, although perhaps at varied rates. In special cases (see Figure \ref{nearlyhalted}) heating may nearly be enough to halt secular cooling, slowing cooling down to a near standstill for 10-100 million years, before more rapid cooling resumes.  

When a planet evolves across a response peak in either direction it experiences a wave of internal heating. Internal temperature changes are in effect tuning the planet's natural frequency to match a constant orbital frequency. In the Maxwell and SAS cases, a single episode of sudden peak heating can occur (regardless of whether the heat is more or less than convection). In the Burgers case two episodes are possible. Changes in orbital forcing, or thermal-orbital feedback could cause planets to evolve though these events multiple times. 

The time it takes a planet to reach equilibrium depends mainly on initial conditions. Our simulations show typical tidal response peaks are crossed rapidly, on the order of 10-50 million years. This will be manifested in the planetary history as a sudden episode of extreme heating, possibly recorded on the planet's surface, followed typically by more moderate equilibrium heat rates. 

We looked for cases where the peak in $W(T)$ could lead to cyclic overshoot events but found the system dynamically overdamped, with cyclic, quasiperiodic, and chaotic solutions prevented by a planet's high thermal inertia and long heat transport timescale. Single overshoots do occur, in particular after heating across a strong resonance peak when the hot stable equilibrium is well below the solidus. 

While equilibrium considerations help limit the range of likely states, uncertainty remains. Extreme equilibrium tidal heating can occur, if by parametric chance the convection curve crosses exactly at the response peak, creating a stable equilibrium very near the maxima in $W(T)$. With favorable parameter choices, such extreme tidal equilibrium solutions can still rise as high as millions of terawatts, as in the fixed $Q$ approach. Therefore, as with the earlier extreme solutions, we need a better understanding of localized partial melting and advective cooling to know if such high outputs occur in real planets.     

\begin{figure*}[t]
\centering
\plotone{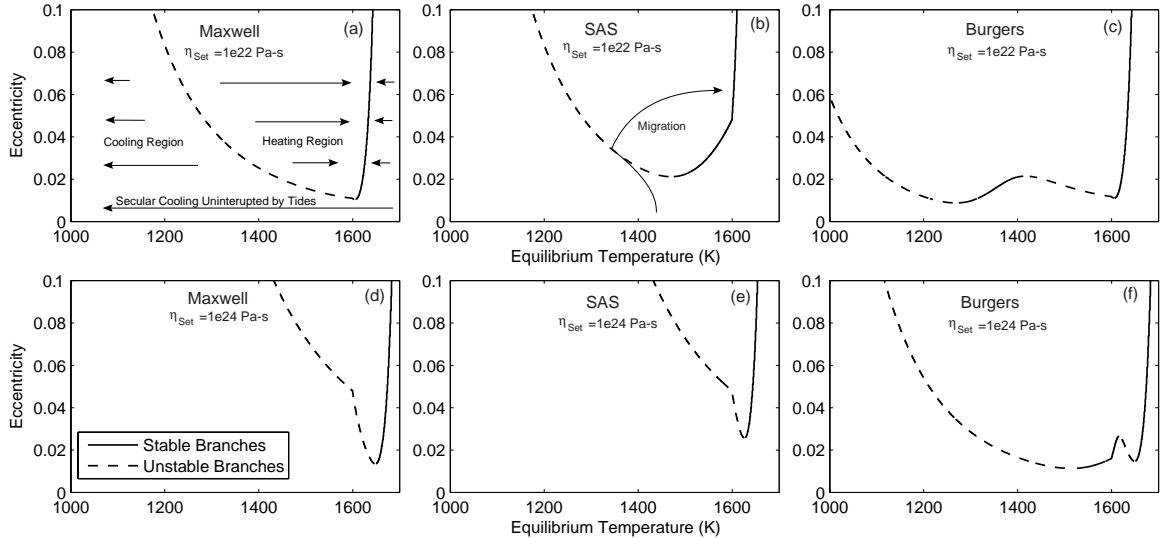}
\caption{Bifurcation diagrams of tidal equilibria for a 1$M_E$ planet in a 16 day orbit, comparing the Maxwell, SAS, and Burgers models, each for two rheologies. Loci of stable equilibria: solid lines. Loci of unstable equilibria: dashed lines. Regions above the equilibria branches represent planetary heating (tidal heat exceeds convection). Regions below the branches represent cooling. Cooling from a high melt fraction will come to a halt on the stable branch if tidal forcing (eccentricity) is high enough. Panel (b) shows a hypothetical trajectory of migrating into a high eccentricity state. Panels (a), (b), (c): baseline rheology. Panels (d), (e), (f): dry or devolatilized rheology with higher defining viscosity, where response peaks shift above the solidus. For some Burgers cases the low temperature stable branch may dominate, particularly at longer periods.}
\label{bifdiagram1}
\end{figure*}

We can generalize behaviors further to incorporate changes in orbital forcing by constructing a bifurcation diagram as shown in Figure \ref{bifdiagram1}. Movement of equilibria points are shown here as functions of eccentricity. Similar diagrams can be created for variations of the semi-major axis. These diagrams are in effect formed by taking a tidal heating curve $W(T)$ and moving it without deforming its shape vertically past the convection curve $\dot{E}_{conv}(T)$. 

Bifurcations occur as the peaks in $W(T)$ make first contact with the convection curve (see Figure \ref{nearlyhalted}a). At low eccentricities, $W(T)$ lies entirely below $\dot{E}_{conv}(T)$, so tidal heating is weak, never competing with cooling, and tidal equilibria are impossible. Bifurcation points represent threshold eccentricities where tidal equilibria come into existence. 

A stable branch, or locus of stable equilibria points, is shown as the solid line. The hot stable branch typically starts near or just after melting and shifts gradually to higher degrees of partial melt. An unstable branch forms on the cold side, shown as the dashed line. Below the bifurcation point, eccentricity is too weak and only secular cooling is possible. 

The region above the two branches represents heating, while the region below represents cooling. Trajectories of planets in the heating region evolve towards the hot stable branch, becoming trapped there at equilibrium. Trajectories evolve away from the unstable branch. Rates of evolution are based on $W(T)$-$\dot{E}_{conv}(T)$. Time varying radionuclides are not included Figure \ref{bifdiagram1}, but will generally add a quasi-stable state at lower temperatures. Orbital migration or eccentricity pumping can move trajectories vertically from the cooling regime into the heating regime or vice versa. Eccentricities in resonances oscillate around a fixed value, which would be represented by a sinusoidal trajectory in this diagram, and has the potential to shift marginal planets between the heating and cooling regimes in a complex manner. 

What these diagrams best demonstrate is that planets will either tend to evolve quickly to the hot stable branch, or will never face tidal heating that can compete with convection and will secularly cool, although perhaps at a slower rate than without tides. This supports our discussion in section \ref{circularization} that maximum dissipation states tend be brief, unless equilibrium occurs at a $W(T)$ peak.

Figure \ref{bifdiagram1}c,f shows the same information for a Burgers body with two response peaks, thus two bifurcation points, two stable branches, and two unstable branches. More complex planetary histories may occur. In particular, planets may become trapped at a colder tidal equilibrium associated with the grain boundary slip mechanism. This cold stable branch is even more pronounced at periods $\geq$20 days and can be the first stable state reached by many planets. However, as inhomogeneous mantles may blur distinct peaks, our Burgers results are best viewed as a demonstration of the increase in behavioral complexity that occurs when additional response frequencies are taken into account.  

\section{Discussion}
\label{discussion}

This work highlights the question of what will be the ultimate shutdown mechanism for an extreme tidal terrestrial planet. Both the fixed $Q$ method and the generally more conservative viscoelastic methods predict that in some circumstances tidal heating can reach millions of terawatts within a planet modeled as homogeneous. Our models of tidal-convective equilibria are very effective in exploring planetary behaviors prior to equilibration, but only coarsely resolve actual equilibrium heat rates due to the assumption of homogeneity.  

\subsection {Inhomogeneous Melting}

Onset partial melting can begin in an inhomogeneous planet at much lower heat rates. To roughly determine the location of melt initiation, we follow \citet{Valencia2006} to calculate the temperature profile with depth, or geotherm $T(z)$, of a tidally heated exoplanet. We estimate the solidus as a function of pressure via the Simon Law using data for forsterite, enstatite, and iron from \citet{Poirier2000} (neglecting high pressure mineral phases and eutectic mixtures). While some models of Earth suggest partial melt already exists at the lithosphere's base, we find the geotherm can begin to intersect the solidus with as little as 30-40 TW of added tidal heat, an approximate doubling of Earth's current output. Where sufficient local partial melting occurs, tidal friction will decrease while continuing in better tuned viscous regions. This suggests how a planet may have difficulty generating the millions of TW solutions found earlier in this paper.
  	
We also find core temperature is a nearly linear function of tidal input, primarily because of the strong linear dependence of the conductive geotherm through the lithosphere on total heat flow. We assume no tidal heat is deposited in the core itself, however small amounts of tidal heat ($\leq$10TW) can shift the geotherm such that the entire core becomes liquid (based on the shallow slope of the Simon Law solidus for pure Iron). Thus even weak tidal exoplanets may have no inner cores, disrupting magnetic dynamo activity, just as with younger exoplanets prior to core crystallization.

Models were tested with an upper-lower mantle thermal separation at 660km depth, as well as asthenospheric only tidal input. Onset melting results were largely the same, except for a weaker dependence of the core temperature on tides, since vigorous upper mantle activity led to thinner conductive lithospheres. Varying the planet's mass, we find above $\sim$2.6$M_E$ the mantle adiabat may curve sufficiently for onset melting to also occur at the core-mantle boundary (using a Birch-Murgnahan equation of state).

As tides increase, we expect partial melt regions to grow in volume, robbing the mantle of material that is properly viscoelastically tuned, and bringing about the process of tidal shutdown. This requires us to abandon a single-geotherm model, since the role and volume fraction of both rising and falling convection plumes becomes key. In the rising plumes, the geotherm may remain above the solidus in much of the upper mantle. In the falling plumes, the geotherm may remain entirely subsolidus. Tidal heat shutdown will be governed by what volume fraction of the mantle exists within higher partial melting zones, and what volume remains best tuned to the orbital period. For certain rheologies, ideal tuning may occur in $\chi$=1-2\% regions, while full detuning occurs at only a few percent greater melt.   

For the most vigorous exoworlds, the assumption of convective heat transfer gives way to a regime dominated by advective heat transport \citep{Spiegelman2001}. Percolation of magma to the surface is vastly more efficient than either bulk convection or conduction, and can move heat rapidly through the bottleneck of a lithosphere. This is akin to the Heat Pipe model suggested \citep{Moore2001, Monnereau2002} to explain Io's apparently strong lithosphere and tall mountains. Introducing advective cooling in Figure \ref{nearlyhalted}a or \ref{burgersbehavior}a is equivalent to the $\dot{E}_{conv}(T)$ curve tipping even more sharply to vertical as soon as melts begin to mobilize. This alone does nothing to set an ultimate limit for tidal heat production. Percolation in fact allows much greater tidal heat rates, since the mantle can stay predominantly rigid, meanwhile running an efficient fluid cooling network, suggesting that tens of millions of terawatt tidal solutions remain geologically possible.    

The true internal extremes of tidal heat production will be governed by models that include grain size and porosity estimates. The amount of cracks and punctures in a lithosphere and the ability of flow networks to endure in a shifting mantle will become vital. An impermeable mantle leads to high melt storage, and a reduction in tidal behavior. An easily permeable mantle leads to an efficient melt network with sparse storage and broad tuned-viscosity zones that allow significant tidal coupling to continue. Vigorous tidal processing may lead to significant global chemical layering between refractory and low melting temperature species.  

\subsection {Magma Oceans}

Hot Earth planets may have insolation supported magma oceans where basal friction due to tidal slosh plays an important role. However mantle tidal heating is unlikely to initiate exposed surface magma oceans from the top down. Fresh lakes of Hawaiian lava solidify to a depth of $\sim$8 meters in one year \citep{Wright1976}. For a heat of fusion near 506 kJ/kg \citep{KojitaniAkaogi1995}, tidal heating can advectively resurface a 1$R_E$ planet at a rate of $\sim$0.0024 cm/yr/TW (given Earthlike background heating). At this rate, 330,000 TW of tidal heat are needed to resurface the entire planet at 8 m/yr, to begin sustained burial of unsolidified material (Assuming all lava flows start at $T_{sol}$ and planetary heat is lost solely through magma volcanism). Full magma ocean build-up via outpourings is likely to require half a million TW or more. So while planet-wide resurfacing may occur, it is difficult for tidal heating to build up a surface magma ocean with no assistance from insolation. Alternatively, sub-lithospheric melting and subsequent thinning may produce insulated near-surface magma oceans or crystalline slush layers at lower tidal rates. 

Individual volcanic vents may produce lava flows greater than 8m/yr and produce significant localized magma lakes. For example, only 370 TW of tidal heat can resurface 1\% of Earth in 1m deep lava flows per year. Thin-layer global resurfacing as on Io is unlikely for viscous lavas. This supports the notion of searching for small radiantly cooled hotspots on supertidal exoplanets.     

Self gravity in $W(T,\omega)$ raises the possibility of a supertidal magma ocean planet. Direct fluid tidal work is very small, due only to the low fluid viscosity. Self-gravity allows for greater peak outputs through resonance by adding an effective $\sim2\times10^{11}$ Pa spring. For a fluid planet:

\begin{equation}
Im(k_2) = -\frac{3}{2} \left( \frac{19\eta_{liq}\omega}{2\beta-(19\eta_{liq}\omega)^2} \right)
\end{equation}

Response timescales for a fluid Earth are on the order of seconds even for the thickest conceivable magmas ($\eta_{liq}$ = 0.01 to 1000 Pa$\cdot$s depending strongly on composition, pressure, and volatile content \citep{McBirney1984}). Still, the climb toward resonance at very short periods does occur in the fluid regime just beyond $T_{brkdwn}$. Fluid tidal heat may play a role for terrestrial planets below 1 day orbits whose mantles never crystallized, or whose insolation induced magma oceans have grown to great depths. 

Following Murray and Dermott, surface deformations of 1-10km are possible on extreme tidal bodies in 1-2 day orbits (deformation is only of the order 300m for Io, and 20m for Europa). This suggests massive faulting and cracking if a lithosphere were to remain brittle, or powerful tidal currents if material is melted. Note $\epsilon_{max}$ is a function of the orbit only, not on $Q$ or $\dot{E}_{tidal}$. The structural geology of extreme periodic fault excitation is particularly interesting in light of the tiger-stripe stress concentrations on Enceladus.   

\subsection {Habitable Zone Modifications}

Finally, we briefly consider tides and habitability. The conventional definition of habitable zone (HZ) uses luminosity to set a range of semi-major axes for stable liquid surface water \citep{Kasting1993} While chemistry and outgassing rates do matter for true habitability, we here consider temperature only. 

\begin{figure}[t]
\centering
\plotone{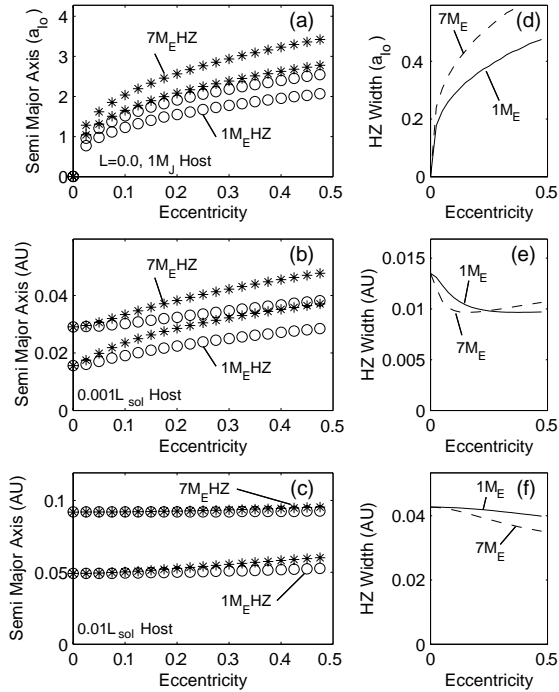}
\caption{
\fontsize{8pt}{10pt}\selectfont
Habitable Zone Modification. Panel (a): Habitable zones for a 1 and 7$M_E$ moon orbiting a nonluminous 1$M_J$ host. Blackbody temperatures for liquid surface water are achieved in orbits similar to the Galilean system with elevated eccentricities.  Panel (b): Habitable zones break up into a spectrum by mass at a M3V dwarf of 0.001$L_{sol}$. Panel (c): Habitable zones at a brighter dwarf of 0.01$L_{sol}$. Here the 1$M_E$ HZ is nearly unaltered by tides, but the super-Earth is large enough that its inner HZ edge is preferentially reduced. All panels use SAS tides with a near-melting isoviscous mantle at $1\times10^{17}$ Pa$\cdot$s, akin to Q=50 $k_2$=0.3, and an Albedo of 0.3. Panels (d), (e), (f): Habitable Zone width reductions as a function of eccentricity for the cases in panels (a), (b), (c). Widths generally contract until tidal heat exceeds insolation heat.}
\label{HZmod1}
\end{figure}

Tidal heating primarily provides additional habitability opportunities for extrasolar moons \citep{Scharf2006}. Tides on larger eccentric moons may not only generate liquid subsurface oceans as with Europa, but also liquid surface water, all at arbitrary distances from a host star. To demonstrate, in Figure \ref{HZmod1}a, we explore a range of orbital parameters for hypothetical 1 and 7$M_E$ moons in orbit around a 1$M_J$ host, using equation \ref{HnSEdot} to estimate tidally driven blackbody surface temperatures. For such large moons, 273K$\leq$$T_{surf}$$\leq$373K is possible for many reasonable values of period and eccentricity given zero contribution from insolation. Large moons of this type may be rare, however we are being conservative looking only at surface temperatures. Subsurface oceans can be sustained at more relaxed masses and eccentricities. Host masses above 1$M_J$ also enhance tides. Thus provided an enduring eccentricity source exists, tides may support island habitable zones around nonluminous primaries such as outer planets, pulsars, ultracool dwarfs \citep{Martin1999}, isolate planets \citep{Osorio2000}, and ejected planets that either retain or accumulate large resonant satellites \citep{DebesSigurdsson2008}. 

A second aspect of tides and habitability is how tidal heat modifies existing habitable zones. Tides can break a HZ into a spectrum by mass, as in Figure \ref{HZmod1}b. In general tides will also reduce HZ widths, due to the $a^{-6}$ term in equation \ref{HnSEdot}. A given increment in eccentricity will lead to a greater change in surface temperature at the inner edge of a HZ than at the outer edge. Width reduction occurs rapidly at low eccentricities, as in Figure \ref{HZmod1}e. Some width recovery occurs once tides become the dominant heat source, but this effect will be rare due to the high eccentricities required. Our calculations do not include the effect of average eccentric insolation, and thus we do not calculate beyond e=0.5. Width reduction is more rapid for super-Earths, and can occur preferentially in special borderline cases as in Figure \ref{HZmod1}f. 

At stars brighter than $\sim$0.01$L_{Sol}$ habitable zones will be unaltered by tides. One can think of a region in semi-major axis space akin to a habitable zone called a tidal zone (TZ), a region where tides can influence surface temperature, typically very close to a star. For G and K stars, the habitable zone is much further out than the tidal zone, and tides never alter habitability. For M3V dwarf stars and dimmer, the HZ and TZ begin to overlap. At this transition, preferential HZ reduction occurs at the inner edge (for 1 vs. 7 $M_E$). At lower luminosities, a HZ is primarily shifted outwards with less width reduction. While dwarf stars such as GJ876 ($0.0124 L_{sol}$) have not been traditional targets for life searches, interest has risen since moons of gas giants can overcome the problem of spin-synchronization, and ice shells or strong magnetic fields can improve radiation shielding. In all cases we have used an SAS planet with a near-solidus isoviscous mantle of $1\times10^{17}$ Pa$\cdot$s, what may be considered ideal viscoelastic tuning. These results are similar to using $Q$=50 and $k_2$=0.3 and thus similar to the measured Earth response values summarized in \citet{KaratoSpetzler1990}. Cooler or devolatilized planets may be too stiff for tides to couple this effectively. 

\section {Conclusions}

In this paper we have shown how a range models produce extreme tidal heating in short period terrestrial exoplanets. The existence of broad regions of extreme tidal solutions lying alongside negligible solutions is robust to parameter uncertainty. However this dual nature makes it difficult to specify a given planetary heat output based on tidal forcing strength alone, without knowledge of the interior. Broadly we find tidal heating in excess of radionuclide heating occurs below approximately 10-30 day orbital periods. Tidal heating in excess of insolation occurs only below $\sim$2 day orbits, and may not be realized due to onset melting and tidal decoupling. Tidal heating around M dwarf stars has the best chance to match or exceed insolation.  

Due to the dominance of insolation at short periods, tidal heating has a negligible impact on the blackbody surface temperature of planets hosted by K class stars and brighter. For such worlds, tidal relevance is predominantly internal. When insolation is strong, moderate temperature elevations due to extreme tides of 1-5 degrees may easily be masked by a small percent uncertainty in albedo. Observational confirmation of extreme tides on a candidate planet suggested by eccentricity should therefore focus primarily on chemical and hotspot signatures. 

Our viscoelastic parametric studies show broad regions in phase space where short period exoplanet tidal heating is both geologically negligible ($\leq$1 TW), and extreme ($\geq$400TW), with a narrow band of moderate tidal cases (1-400 TW) separating the regions. The Burgers body and SAS models can reveal otherwise hidden features in tidal evolution not manifested by the simpler Maxwell model, including extra episodes of warming and extra equilibrium points. Use of the creep compliance term $\delta\!J$ in place of the bulk compliance $J$ can lead to significant variations in results. The distinct response peaks that occur in a Burgers model with well defined parameters may however be blurred in a real planet with a range of heterogeneities.  

If a planet can ever couple significantly with tides, it will either secularly cool into a stable equilibrium with convection, or undergo 1-2 rapid warming episodes before stabilizing. Equilibrium heat outputs are sensitive to model choice, can be negligible or extreme, and can occur both above or below the solidus. Tidal heating typically reaches equilibrium with convection in a few million years. Stable equilibria can be shifted, created, or destroyed by changes in eccentricity.  

Melt advection is the key to determining the limits of terrestrial tidal heat production. True tidal equilibria must be derived from a balance between partial melt production regions and melt percolation rates, invoking separate adiabats for rising and falling convective plumes. Tides are unlikely to generate surface magma oceans on their own, but due to high insolation at short periods, they may help control the depth of preexisting magma oceans. 
 
Extreme tides, if present, will mainly alter the habitable zones for short period planets of lower luminosity M dwarf stars. Habitable zones may be shifted outwards, preferentially reduced from their inner edges, and partly or completely split into a spectrum by mass. Tides at brighter stars will only influence uninhabitable hot worlds, or may alter cold resonant moon systems at arbitrary distances from the host star. 

\section{Acknowledgments}

This work has been supported by NASA Grant NNG05GP17, NSF Grant EAR0440017 and the Origins of Life Initiative at Harvard University. We wish to thank Daniel C. Fabrycky, Jack J. Lissauer, Ruth Murray-Clay, and Sarah T. Stewart-Mukhopadhyay for many helpful reviews and comments. We especially thank Dr. Michael Efroimsky for his uniquely careful, detailed, and thoughtful comments on our work.

\bibliography{henningetal2009}
\bibliographystyle{apj}

\end{document}